# Does The Pioneer Anomalous Acceleration Really Exist?

Walter Petry
Mathematisches Institut der Universitaet Duesseldorf, D-40225 Duesseldorf
E-mail: wpetry@meduse.de
petryw@uni-duesseldorf.de

Abstract: The analysis of the Pioneer 10 and 11 data demonstrated the presence of an anomalous Doppler frequency blue-shift drift which is interpreted as an anomalous acceleration. The Doppler frequency dirft follows by considering the motions of the Pioneers in the universe, i.e. it is of cosmological origin. There is no anomalous acceleration.

## 1. Introduction

There are many papers confirming an anomalous Doppler frequency blue-shift drift of the Pioneers (see e.g. Anderson et al. [1], Markwardt [2], Truyshev et al. [3], etc.). Generally it is interpreted as an anomalous acceleration of the spacecrafts. But it is difficult to explain this anomalous acceleration with standard physics.
In this paper it is shown that the motion of the spacecrafts must be studied in the universe. The Doppler frequency drift follows by a nonlinear variation of time in the universe for the observer, i.e. it is of cosmological origin and there is no anomalous acceleration of the spacecrafts.

## 2. Summary of some Results

Let us start from a theory of gravitation in flat space-time studied in several papers (see e.g. [4]). A summary of the theory with applications can be found in paper [5] where references to the detailed studies are stated.
Subsequently, we summarize some results of paper [6] which are used in the following.
The flat space-time theory of gravitation has a flat space-time background metric

$$(ds)^2 = -\eta_{\alpha\beta} dx^\alpha dx^\beta. \qquad (2.1)$$

The gravitational field is desrcibed by symmetrc tensor $g_{ij}$ satisfying covariant (with respect to the flat space-time metric (2.1)) differential equations of order two where the source of the field is the total energy-momentum tensor inclusive the gravitational field. The proper time (atomic time) is defined by

$$c^2 (d\tau)^2 = -g_{\alpha\beta} dx^\alpha dx^\beta. \qquad (2.2)$$

The application of the theory to homogeneous, isotropic, cosmological models starts with the pseudo-Euclidean geometry, i.e.

$$(\eta_{ij}) = diag(1,1,1,-1) \qquad (2.3)$$

where $x^1, x^2, x^3$ are Cartesian coordinates and $x^4 = ct$. The four-velocity of the universe is

$$u^i = 0 \ (i=1,2,3), \quad u^4 = c\frac{dt}{d\tau} \qquad (2.4)$$

and the potentials are

$$(g_{ij}) = diag(a^2(t), a^2(t), a^2(t), -1/h(t)) \qquad (2.5)$$

where $a(t)$ and $h(t)$ satisfy two coupled differential equations of order two with the initial conditions at present time $t_0 = 0$

$$a(0) = h(0) = 1, \quad \dot{a}(0) = H_0, \quad \dot{h}(0) = \dot{h}_0. \qquad (2.6)$$

Here, the dot denotes the time-derivative, $H_0$ the Hubble constant and $\dot{h}_0$ is a further constant of integration which is zero for Einstein´s theory. It follows that there are non-singular, cosmological models under natural conditions in contrast to Einstein´s theory. The functions $a(t)$ and $h(t)$ must not be known for the subsequent considerations but they can be found in papers of the author.

The Newtonian approximation of a perfect fluid in this universe is stated. Let

$$\rho(x,t), \ v(x,t) = (v^1(x,t), v^2(x,t), v^3(x,t)) \qquad (2.7)$$

be the density, resp. the three-velocity of the perfect fluid. Then, the density

$$\rho^* = \rho \frac{dt}{d\tau} \qquad (2.8)$$

implies the conserved mass

$$M = \int \rho^*(x',t) d^3 x' \qquad (2.9)$$

and the equations of motion to Newtonian approximation in the universe are

$$\frac{\partial}{\partial t}\left(a^2 \sqrt{h} v^i\right) + \sum_{\alpha=1}^{3} v^\alpha \frac{\partial}{\partial x^\alpha}\left(a^2 \sqrt{h} v^i\right) = -\frac{1}{a\sqrt{h}} k \int \rho^*(x',t) \frac{x^i - x^{i'}}{|x - x'|^3} d^3 x'. \qquad (2.10)$$

Here, $|\cdot|$ denotes the Euclidean norm and $k$ is the gravitational constant.

Assume that a distant atom in the universe is moving with velocity $(v,0,0)$ and emits at time $t_e$ a photon moving to the observer. Put

$$\gamma = \left(1 - \left(a(t_e)\sqrt{h(t_e)}\frac{v}{c}\right)^2\right)^{-1/2} \qquad (2.11)$$

and let $E_0$ be the energy of the photon emitted from the same atom at rest. Then, the observer receives from the moving atom the emitted photon with the energy

$$E(0,t_e) = \gamma^{-1}\left(1 + a(t_e)\sqrt{h(t_e)}\frac{v}{c}\right)^{-1} a(t_e) E_0. \qquad (2.12)$$

### 3. Explanation of the Doppler Frequency Drift

In paper [6] an explanation of an anomalous acceleration of spacecrafts has been given by the use of equation (2.10). But the authors of the papers [1-3] measure a Doppler frequency drift which is interpreted as an anomalous acceleration. Furthermore, an observer uses not the time $t$ defined by the pseudo-Euclidean background geometry (2.3). This follows from the following considerations. Assume that a light ray at an object with distance $r$ from the observer is emitted at time $t_e$ and arrives at time $t_a$ at the observer. It follows from (2.2) by the use of (2.5)

$$r = c \int_{t_a}^{t_e} 1/(a\sqrt{h})(t) dt. \qquad (3.1)$$

Let us consider the relation (3.1) for two different light rays emitted at $t_{e_1}$ resp. $t_{e_2}$ with the arrival time $t_{a_1}$ resp. $t_{a_2}$ then

$$\int_{t_{a_1}}^{t_{e_1}} 1/(a\sqrt{h})(t) dt = \int_{t_{a_2}}^{t_{e_2}} 1/(a\sqrt{h})(t) dt$$

implying

$$\int_{t_{a_1}}^{t_{a_2}} 1/(a\sqrt{h})(t) dt = \int_{t_{e_1}}^{t_{e_2}} 1/(a\sqrt{h})(t) dt. \qquad (3.2)$$

Put $dt = t_{e_2} - t_{e_1}$ and $dt' = t_{a_2} - t_{a_1}$ then (3.2) implies by the use of (2.6)

$$dt' = \frac{dt}{a(t)\sqrt{h(t)}}. \qquad (3.3)$$

Relation (3.3) gives the connection between the time epoch $dt$ at a distant object and the corresponding one $dt'$ as measured by the observer. The whole time $t'$ of the observer since the beginning of the universe till the time $t$ of the pseudo-Euclidean geometry as background geometry is given by

$$t' = \int_{-\infty}^{t} dt/(a\sqrt{h})(t).\tag{3.4}$$

Relation (3.4) defines a unique correspondance between the times $t$ and $t'$. Subsequently, for any time $t$ the corresponding time of the observer is denoted by $t'$ and conversely.

The observer's time (3.3) implies for the relation (2.2) with (2.5) the proper time

$$c^2(d\tau)^2 = -a^2(t)\left(\sum_{\alpha=1}^{3}(dx^\alpha)^2 - (dct')^2\right)\tag{3.5}$$

and for (2.1) with (2.3) the background metric

$$(ds)^2 = -\left(\sum_{\alpha=1}^{3}(dx^\alpha)^2 - (a^2h)(t)(dct')^2\right)\tag{3.6}$$

where $t$ must be replaced by $t'$ by the use of (3.4).

Relation (3.5) implies for the observer that the light velocity in the universe is always the vacuum light velocity. The velocity $v^i(t)$ with repect to the system time $t$ of the pseudo-Euclidean geometry is transformed by the use of (3.3) to the observer's velocity

$$v^{i'}(t') = v^i(t)\frac{dt}{dt'} = (a\sqrt{h})(t)v^i(t).\tag{3.7}$$

Then, the equations of motion (2.10) in the universe have the form

$$\frac{\partial}{\partial t'}(av^{i'}) + \sum_{\alpha=1}^{3}v^{\alpha'}\frac{\partial}{\partial x^\alpha}(av^{i'}) = -k\int \rho^*(x',t(t'))\frac{x^i - x^{i'}}{|x-x'|^3}d^3x'.\tag{3.8}$$

These equations give in analogy to the considerations of paper [6] for several point masses $M_j$ in the solar system by the use of (2.6)

$$\frac{dv_l'}{dt'} = -H_0 v_l' - k\sum_{j\neq l}\frac{M_j(x_l - x_j)}{|x_l - x_j|^3}\tag{3.9}$$

where $v_l'$ denotes the velocity vector of particle $l$. Hence, the observer can not measure an anomalous acceleration in the universe because the first expression on the right hand side of (3.9) is too small.

Let us now consider the energy of a photon emitted at time $t_e$ from an atom moving in the universe away from the observer. Formula (2.12) gives by the use of (3.7) to the first order in the velocity by the use of $E = h\nu$ ($h$: Planck constant) for the frequency of the arriving photon

$$\nu(0,t_e) \approx \left(1 - \frac{v'(t_e')}{c}\right)a(t_e)\nu_0\tag{3.10}$$

where $\nu_0$ is the frequency emitted at present by the same atom at rest (reference frequency). In the general case where the velocity vector and the line of sight enclose an angle $\vartheta$ the velocity $v'$ must be multiplied with $\cos\vartheta$.

The total Doppler frequency drift is given by

$$\frac{d}{dt_e'}(\nu_{obs}(t_e') - \nu(0,t_e)) = \frac{d}{dt_e'}\left(\nu_{obs}(t_e') - \left(1 - \frac{v'(t_e')}{c}\right)\nu_0\right) + \frac{d}{dt_e'}\left(\left(1 - \frac{v'(t_e')}{c}\right)(1 - a(t_e))\nu_0\right)\tag{3.11}$$

where $\nu_{obs}$ is the observed frequency drift. The first expression on the right hand side is the measured frequency drift (see e.. [3])

$$\dot\nu \approx 6\cdot 10^{-9}\, Hz/s.\tag{3.12}$$

The last expression of equation (3.11) is given by Taylor expansion and the use of (2.6)

$$\frac{d}{dt_e{}'}\left(\left(1-\frac{v'(t_e{}')}{c}\right)(1-a(t_e))\nu_0\right) \approx (a\sqrt{h})(t_e)\frac{d}{dt_e}\left(-H_0 t_e \nu_0\right) \approx -H_0\nu_0. \qquad (3.13)$$

The used reference frequency (see [1]) is

$$\nu_0 = 2.29 \cdot 10^9 \, Hz \qquad (3.14)$$

and presently assumed best Hubble constant

$$H_0 \approx 70\frac{km}{\sec Mpc} \approx 2.3 \cdot 10^{-18}\frac{1}{s} \qquad (3.15)$$

give for the second expression of formula (3.11)

$$-H_0\nu_0 \approx -5.3 \cdot 10^{-9}. \qquad (3.16)$$

Hence, the total Doppler frequency drift (3.11) in the universe is by the use of (3.12) and (3.13) with (3.16) about zero. Therefore, the measured frequency drift does not imply an anomalous acceleration but follows by considering the energy of the photons in the universe, i.e. it is of cosmological origin.

**References**
[1] J.D. Anderson, P.A. Laing, E.L. Lau, A.S. Liu, M.M. Nieto and S.G. Turyshev, Phys.Ref. D**65**,082004(2002).
[2] C.B. Markwardt, arXiv: gr-qc 0208046.
[3] S.G. Turyshev, M.M. Nieto and J.D. Anderson, arXiv: gr-qc/0503021.
[4] W. Petry, Gen.Rel.Grav.**13**, 865 (1981).
[5] W. Petry, in: Recent Advances in Relativity Theory, Vo. II (Eds. M.C. Duffy and M. Wegener) Hadronic Press, Palm Harbor FL 2002, p.196.
[6] W. Petry, Z. Naturforsch. **60a**, 255 (2005).